\documentclass[letterpaper]{JHEP3}
\usepackage{graphics}

\newcommand{\fft}[2]{{\frac{#1}{#2}}}

\title{The cosmic role of tachyon in the type 0 strings}

\author{Wen-Yu Wen\\
Physics department \\
National Taiwan University\\
Taipei, Taiwan, R.O.C.\\
\email{steve.wen@gmail.com}}

\abstract{We present a new class of solution to the
ten-dimensional type 0 effective action. Given a generic potential
of tachyon field, there exist phases where tachyon is either
frozen at local extremals or free to propagate along flat
directions.  In the latter phase, a cosmology model is proposed
where the tachyon plays the role of time.}

\begin{document}

\section{Introduction}

Ever since the AdS/CFT correspondence is
formulated\cite{Maldacena:1997re,Witten:1998qj}, many supporting
evidences came from those BPS states where at least some fraction
of supersymmetry is preserved.  Out of them, the case of type IIB
strings in the background $AdS_5\times S^5$ was mostly
studied\cite{Maldacena:1997re}.  The vanishing beta function of
conformal field theory, which origins from non-dilatonic D3 brane,
strongly suggests a possible extrapolation from the strong-coupled
regime to the weakly-coupled one or vice versa.  While the former
is calculable in the classical supergravity, the latter is done by
perturbative field theory. Such correspondence has been
generalized to the one between non-conformal field theory and
dilatonic branes\cite{Itzhaki:1998dd,Boonstra:1998mp}.  There
exist half BPS domain-wall solutions, in which live the dual super
Yang-Mills theories whose effective coupling runs with non-trivial
dilaton background . An important feature associate with running
coupling in such non-conformal theories is that the radial
direction of AdS space plays the role of energy scale.  Pushing a
tested brane from the boundary into the core of AdS space
corresponds to the induced field theory running from its UV
towards a IR point.  Though, of course, this generalization is at
expense of losing any advantage from conformal theory, it
approaches closer to realistic models of our world.

Another essential ingredient in model building, especially for
cosmology, is the time-dependent solution.  It has been a
challenge in this braneworld scenario to construct time-dependent
solutions fit into the description of modern
cosmology\cite{Khoury:2001bz}. One proposal towards this direction
is to understand the big bang singularity in the context of
null-like cosmology\cite{Craps:2005wd}.  Along this line of
thinking, authors in \cite{Chu:2006pa,Lin:2006ie} constructed a
class of quarter BPS solution, which can be seen as the pp-wave in
the AdS space with non-trivial dilaton background.  As shown in
\cite{Lin:2006ie} that in general the effective coupling not only
runs with the radial direction (energy scale) but also the
light-cone {\sl time} on the domain-wall.  Therefore a big bang
cosmology can be proposed on the boundary and its evolution
depends on the profile of scalar field, or equivalently the
warping factor of boundary metric.

In this note, we would like to generalize the solutions found in
\cite{Lin:2006ie} in the context of non-supersymmetric type $0$
strings.  One might wonder how can supersymmetric solutions fit
into non-supersymmetric theories, and worry if the existence of
tachyon field signals the instability of solutions.  However, we
will show that one can easily associate solutions in the type II
strings with type $0$ once the tachyon field is either frozen or
in the null-like background. As a toy model, we also investigate
the possibility that the role of comic time is played by the
null-like tachyon.  We also remark that cosmology for a D$3$-brane
probing in the type $0$ tachyonic background has been studied in
\cite{Papantonopoulos:2006eg}.

This note is organized as follows:  we start with a very brief
introduction to type $0$ strings and present the effective action
in section $2$.  Then in section $3$ we will discuss the solutions
with frozen tachyon and in section $4$ with propagate one. A
cosmology model is proposed in section $5$ and discussion in
section $6$.

\section{Type 0 strings and the effective action}

We summarize some basic features of type $0$ strings and their
gravity effective
action\cite{Polyakov:1998ju,Dixon:1986iz,Seiberg:1986by,Klebanov}.
Type $0$ strings are purely bosonic strings, obtained from the
closed string by a diagonal GSO projection which excludes
space-time fermions.  The field contents include a tachyon field
from the $(NS-,NS-)$ sector and doubled set of D-branes coupling
to form fields in the $(R\pm,R\pm)$ sectors.  There is a closed
string tachyon that renders the Minkowski vacuum unstable,
nevertheless this does not prohibit the existence of other
possible vacua where negative lower bound for the mass square are
allowed, such as the AdS space. It was also pointed out in
\cite{Klebanov} that the background RR-flux provides a positive
shift to the mass square of the tachyon, thus the tachyonic
instability can be cured for sufficiently large flux.  In order to
obtain the effective action, one observes that all the tree level
amplitudes which involve only the fields from the $(NS+,NS+)$ and
$(R\pm,R\pm)$ sectors are identical to those in type II theory,
while the worldsheet supersymmetry constrints the $(NS-,NS-)$
sector of the action.  It turns out that the effective action
takes the following form\cite{Klebanov}:
\begin{equation}\label{effect_act}
e^{-1}{\cal L} = R - \fft12(\partial \phi)^2 - \fft12 (\partial
T)^2 - V(T) e^{\fft12\phi} - \fft{f(T)}{2(p+2)!}e^{\kappa\phi}
F^2_{(p+2)},
\end{equation}
where $\kappa=\fft14(d-2p-4)$.
The perturbative analysis suggests $V(T)=(d-10)-\fft{d-2}{8}T^2 +
{\cal O}(T^4)$ an even function of $T$, while $f(T)=1+T+\fft12
T^2$ incorporates kinetic term of $F$ with three-point and
four-point interaction of $TFF$ and $TTFF$\cite{Martelli}. In
\cite{Klebanov} the authors considered a stack of
3-brane\footnote{Due to the doubled RR fields, D$3$ brane in type
0 can be either electric- or magnetic-charged. $5$-form field
strength is no longer self-dual and thus $F^2$ term will
contribute to the equations of motion. Later we will consider this
special case.} in the critical dimension. In the presence of RR
field, the tachyon condenses near the extremum of $f(T)$, i.e.
$T=-1$ and its effective mass square is shifted to a positive
value. The running dilaton thus has a source coming from condensed
tachyon and the conformal invariance of dual gauge theory is
broken.  It was also argued this theory has the RG flow from this
UV point to a IR point (infinite coupling) at $T=0$, although
explicit solutions along the flow are still unknown.

\section{Vacuum solutions with frozen tachyon}

Here we also consider the critical strings, i.e. $d=10$, but an
arbitrary $V(T)$ with at least one local vacuum around $T_0$ where
$V(T_0)=V'(T_0)=0$. Besides, in stead of restricting $f(T)$ only
up to $T^2$ term, we allow $f(T)=e^T$ to incorporate interaction
of arbitrary number of tachyon with RR form field, i.e. terms like
$T^nFF$ for arbitrary $n$ are considered.  We would like to show,
after redefinition of some fields, that a new class of solution is
also consistent with tachyon condensation. This new class of
solution depends not only on the radial coordinate $r$, but also
on the light-cone coordinate $u$, which can be seen as a
generalization of the solution in \cite{Lin:2006ie}

For $\kappa \neq 0$\footnote{For $\kappa =0$, i.e. $p=3$, the
following redefinition is not working and we have to stick to the
set of equations \ref{eom_old}.  We will separately discuss this
issue later.}, we may redefine $T$ and $\phi$ in the action
\ref{effect_act}:
\begin{eqnarray}
&&\tilde{\phi}:= \phi+ \fft{1}{\kappa}\ln{f}\\
&&\partial\tilde{T} := \sqrt{1-\kappa^{-1}(\fft{f'}{f})^2}\partial
T\\
&&\tilde{V} := f^{-\fft{1}{2\kappa}}V
\end{eqnarray}
For our assumption $f(T)=e^T$, $\partial T$ can be further
integrated as $\tilde{T}=\sqrt{1-\kappa^{-1}}T+c$, where $c$ is an
integrated constant.

After redefinition, the action becomes
\begin{equation}
e^{-1}\tilde{\cal L}= R - \fft12(\partial\tilde{\phi})^2
-\fft12(\partial \tilde{T})^2 -
\tilde{V}e^{\fft12\tilde{\phi}}-\fft{1}{2(p+1)!}e^{\kappa\tilde{\phi}}F^2
\end{equation}
Notice that the tachyon, after redefinition, does not explicitly
couple to RR field. Then the equations of motion become:
\begin{eqnarray}
&&R_{\mu\nu}=\fft12
\partial_{\mu}\tilde{\phi}\partial_{\nu}\tilde{\phi} + \fft12
\partial_{\mu}\tilde{T}\partial_{\nu}\tilde{T} +
\fft18g_{\mu\nu}\tilde{V}(\tilde{T})e^{\fft12\tilde{\phi}}+e^{\kappa\tilde{\phi}}T_{\mu\nu}\label{eom_metric}\\
&&\nabla_{\mu}(e^{\kappa\tilde{\phi}}F^{\mu\cdots})=0\nonumber\\
&&\nabla^2\tilde{\phi} = \fft12
\tilde{V}(\tilde{T})e^{\fft12\tilde{\phi}} + \fft{\kappa
}{2(p+2)!}e^{\kappa \tilde{\phi}} F^2 \label{eom_phi}\\
&&\nabla^2\tilde{T} =
\tilde{V}'(\tilde{T})e^{\fft12\tilde{\phi}}\label{eom_tachyon}
\end{eqnarray}

At the first glance, this set of equations of motion are almost
same as those of the bosonic fields in \cite{Lin:2006ie} except
tachyon is absent there for the type II strings.  This may not be
a surprise because both string theories share the same worldsheet
supersymmetry, though different at GSO projection to the spacetime
supersymmetry.  Hence it is quite straightforward to apply
solutions from \cite{Lin:2006ie} but discard all their
supersymmetric properties. Indeed, we have found the following
solution for constant $\tilde{T_0}$ at some vacuum where
$\tilde{V}(\tilde{T}_0)=\tilde{V}'(\tilde{T}_0)=0$.
\begin{eqnarray}\label{ansatz}
&&ds^2=r^{(\kappa+2)^2/2}a(u)^2b(u)^{-\kappa/(2(\kappa+2))}(-2dudv+h(u,r,\vec{x})du^2+d\vec{x}^2_{(p-1)})\\
&&\qquad\qquad
+r^{\kappa^2/2}b(u)^{-\kappa/(2(\kappa+2))}(r^{-2}dr^2+d\Omega^2_{(8-p)}),\\
&&e^{\tilde{\phi}}=r^{-\kappa(\kappa+2)}b(u),\\
&&F_{uv\cdots pr}= 2(\kappa+2)r^{2\kappa+3}, \\
&&\tilde{T}=T_0,
\end{eqnarray}
where $b(u)=(a(u))^{\fft{4-\kappa^2}{\kappa}}$.  Here we use the
same notation as in\cite{Lin:2006ie}.  $u,v$, $\vec{x}$ and radial
$r$ are coordinates in $AdS$, in which $u,v$ are light-like. This
solution describes a $AdS_{p+2}\times S^{8-p}$
geometry\footnote{As shown in \cite{Lin:2006ie} that $AdS\times S$
geometry is only manifested in the dual frame $g^D$, which is
related to the Einstein frame $g^E$ by a Weyl transformation
$g^D=e^{\kappa\tilde{\phi}/2(\kappa+2)}g^E$.} in the pp-wave
background coupled to the dilaton $\tilde{\phi}$, the tachyon
$\tilde{T}$ and the RR form field $F_{(p+2)}$.

\section{Vacuum solutions with propagate tachyon}

Other than the previous static solution with constant tachyon, we
are actually allowed to switch on $u$-dependence of tachyon field
if there is a flat direction in the vacuum, in which tachyon is
free to propagate.  In such a case, equation \ref{eom_tachyon} is
satisfied due to the null-like Killing direction along $u$, i.e.
$g^{uu}=0$. We are left with equation \ref{eom_metric},
$uu$-component in particular. It turns out that $\tilde{T}$ is
only constrained by the following equation,
\begin{equation}\label{constraint}
(\partial_u\tilde{T})^2 =  4\partial_u^2 \ln a -\fft{16}{\kappa^2}
(\partial_u\ln a)^2 - \vec{\nabla}^2 h +
r^{(5-p)}a^2[(p-8)r\partial_rh - r^2\partial_r^2h].
\end{equation}
As shown in \cite{Lin:2006ie}, $h(r,u,\vec{x})$, and thus $a(u)$,
is determined by two arbitrary functions $P(u)$ and Q(u) in the
following way:
\begin{equation}
h(u,r,\vec{x})= {2 P(u) \over
(p-5)}r^{(p-5)}+h_0(u,\vec{x})r^{(p-7)}+h_1(u,\vec{x}),
\end{equation}
where $\vec{\nabla}^2 h_0(u,\vec{x})=0$ and $\vec{\nabla}^2
h_1(u,\vec{x})=4 Q(u)$.  Therefore the profile of $\tilde{T}(u)$
serves as another arbitrary function to determine $a(u)$.  On the
other hand, if $\tilde{T}(u)$ is invertible, i.e.
$u=u(\tilde{T})$, then equation \ref{constraint} tells us how does
warping factor $a(u[\tilde{T}])$ evolves with tachyon $\tilde{T}$.
We will elaborate this point with some examples in the next
section.

\section{Boundary cosmology}
As shown in \cite{Lin:2006ie}, this class of solution in the
context of type II strings corresponds to the near horizon region
of D$p$ brane where the pp-wave travels in the world volume, hence
$8$ supercharges are preserved.  In the context of type $0$
strings, although there is no supersymmetry, it still corresponds
to a similar setting except that there are two kinds of branes
coupled to two different RR sectors, which are not distinguished
in the solution.  The coupling running with light-cone coordinate
$u$ mimics a cosmology on the brane world volume (or domain wall
upon compactification of transverse sphere).  In order to
associate with boundary cosmology, one identify $u$ as the {\sl
conformal} time and introduce a {\sl comoving} time $dt=a(u)du$,
then equation \ref{constraint} can be rewritten in terms of $P,Q$
as Friedmann-like equation:
\begin{equation}\label{T_profile}
-\fft{\dot{\tilde{T}}^2}{4} + \fft{\ddot{a}}{a} -
\fft{4}{\kappa^2}H^2 = \fft{Q}{a^2} + P,
\end{equation}
where the dot denotes the derivative with respect to $t$, and {\sl
Hubble parameter} $H:=\fft{\dot{a}}{a}$. In the phase of constant
tachyon, we have exactly the same discussion as in
\cite{Lin:2006ie} given the tachyon's degree of freedom is frozen.
In summary, vanishing $P,Q$ will render a flat universe where
$a(t)$ as a monotoneous function of time, nonzero $Q$ signals an
open or closed universe, and function $P$ encodes the information
of density and pressure.

In the phase of propagate tachyon, this extra degree of freedom
contributes to equation \ref{constraint} in term of tachyon's
kinetic energy.  Out of many possible profile of $\tilde{T}(u)$,
we pay attention to just one simple but interesting example -
linear tachyon, say $\tilde{T}\sim 2\omega u$.  We notice that
with this linear profile the tachyon kinetic term behaves like a
constant $Q$-term.  This brings us the following possible
universes, classified by the ratio $\gamma=Q/\omega^2$:

\subsection*{a.$\quad \gamma=-1, P=0$}

\begin{figure}
\center{
\includegraphics{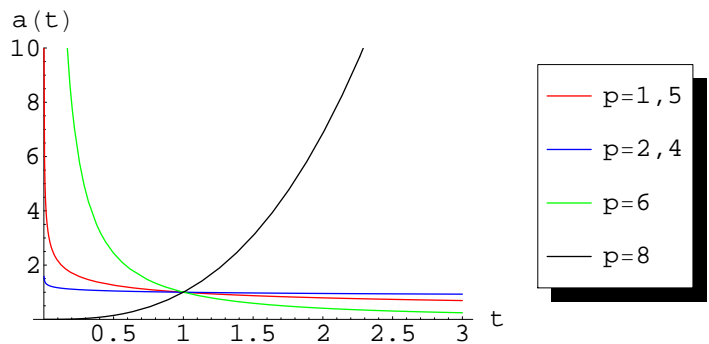}}
\caption{Plot of $a(t)$ for different D$p$-brane, where the
boundary cosmology is a flat vacuum, i.e. $\gamma=-1,P=0$.}
\label{fig1}
\end{figure}

This describes a $\sl flat$ vacuum universe, where the warping
factor scales like
\begin{equation}
a(t)\sim t^{\alpha},\qquad \alpha=(1-4/\kappa^2)^{-1},
\end{equation}
or equivalently $a\sim t^{Ht}$.  We summarize results for various
D$p$-branes in the figure \ref{fig1}.

\subsection*{b.$\quad \gamma=-1, P \propto a^{-p}$}

\begin{figure}
\center{
\includegraphics{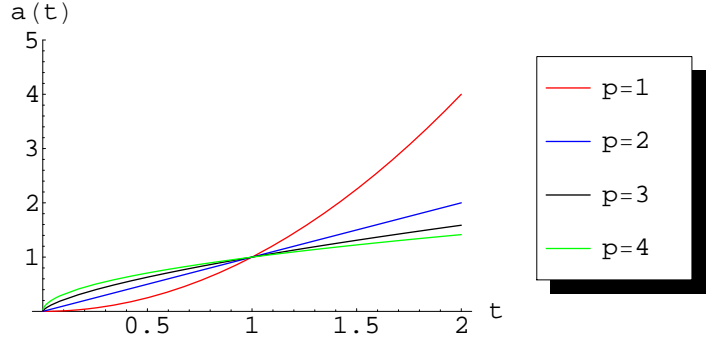}}
\caption{Plot of $a(t)$ for some D$p$-branes, where the boundary
cosmology is in the matter-dominated phase, i.e.
$\gamma=-1,P\propto a^{-p}$.} \label{fig2}
\end{figure}

This describes a flat $\sl matter$-dominated universe, where
\begin{equation}
a(t) \sim t^{\fft{2}{p}}.
\end{equation}
We sample results for some D$p$-branes in the figure \ref{fig2}.

\subsection*{c. $\quad \gamma=-1, P \propto a^{-p-1}$}

This describes a flat $\sl radiation$-dominated universe, where
\begin{equation}
a(t) \sim t^{\fft{2}{p+1}}.
\end{equation}

\subsection*{d. $\quad \gamma>-1, P=0$}

This describes a $closed$ vacuum universe, where
\begin{equation}\label{universe_closed}
a(T) \sim c_1\cos{\fft{|\kappa|\sqrt{1+\gamma}}{4}T} +
c_2\sin{\fft{|\kappa|\sqrt{1+\gamma}}{4}T},
\end{equation}
where $c_1$ and $c_2$ are integral constants determined by the
initial condition of universe.  It is more convenient here to
express $a(t[T])$ as a function of the tachyon field.  The period
of each cycle is given by $4\pi/(|\kappa|\sqrt{1+\gamma})$.

\subsection*{e. $\quad \gamma<-1, P=0$}

This describes an $open$ vacuum universe, where
\begin{equation}\label{universe_open}
a(T) \sim d_1\exp{\fft{|\kappa|\sqrt{-\gamma-1}}{4}T} +
d_2\exp{-\fft{|\kappa|\sqrt{-\gamma-1}}{4}T},
\end{equation}
where $d_1$ and $d_2$ are integral constants determined by the
initial condition of universe.  Though the analysis of
nonvanishing $P$-term while $\gamma<-1$ is skipped for simplicity,
we would like to point out that a constant of $P$-term like can be
generated by taking $\tilde{T}\propto \int{a(u)du}$.

\section{Discussion}

In this section, we have several comments on the solutions
obtained above.  At first we argue the effective coupling would
have the same behavior as in \cite{Lin:2006ie}.  Secondly we
comment on the case of decoupled dilaton, i.e. $\kappa=3$.  At
last, we would like to promote a little the idea of null-like
tachyon as comic time.

\subsection*{Dual Yang-Mills and the effective coupling}

In the spirit of AdS/CFT and its generalization Domain-Wall/QFT,
it is quite interesting to see what is the effect to the dual
Yang-Mills coupling by introducing tachyon field in the type $0$
strings\cite{Klebanov}.  In our redefinition of field, however,
the tachyon coupled to RR field has been absorbed into dilaton
field $\tilde{\phi}$, thus the effective coupling in either phase
mentioned above should be the same as it was found in
\cite{Lin:2006ie}.  We recall that the effective $SU(N)$
Yang-Mills coupling in the world volume is
\begin{equation}
g^2_{eff} \sim Ne^{\tilde{\phi}}=[g^2_{YM}
Nr^{-2\kappa}]^{\kappa+2\over 2}, \qquad g_{YM}:=\bar{g}_{YM}
(b(u))^{1/(\kappa+2)}.
\end{equation}
The readers are directed to \cite{Lin:2006ie} for detail
discussion.

\subsection*{The $\kappa=0$ vaccum}

We would like to discuss, for $\kappa=0$, solutions to
(\ref{effect_act}) at tachyon vacua $V(T)=V'(T)=0$.  Since in
generic $f'(T)\neq 0$ for our choice of $f(T)$, we do not have
solutions with constant tachyon as before. In fact $T$ has to be a
function of $r$ and maybe $u$ as well, and equations of motion
translates into
\begin{eqnarray}
&&R_{\mu\nu}=\fft12 \partial_{\mu}\phi\partial_{\nu}\phi + \fft12
\partial_{\mu}T\partial_{\nu}T -\fft14 g_{\mu\nu}e^{-T}Q^2\\
&&\sqrt{-g}f(T)F^{uv\cdots}=Q\\
&&\nabla^2\phi = 0\\
&&\nabla^2T = -\fft12e^{-T}Q^2\label{toda}
\end{eqnarray}
$Q$ is a conserved charge for tachyon-coupled RR field.  Notice
that the tachyon field $T$ obeys the Toda equation in one higher
dimension.  To see this we add one {\sl virtual} dimension $z$ and
set $\hat{T}(x^{\mu},z)=T(x^{\mu})+\fft{Q}{\sqrt{2}}z$, then
equation \ref{toda} becomes
\begin{equation}
\nabla^2\hat{T}+\partial_z^2e^{-\hat{T}}=0
\end{equation}
The dilaton $\phi$ could be either a constant or a function of $u$
if $g^{uu}=0$ is satisfied again. It would be interesting to see
if a deformation of metric \ref{ansatz} exists for the above
equations of motion.

\subsection*{Null-like tachyon as comic time}
In equations \ref{universe_closed} and \ref{universe_open}, it
suggests that the tachyon field plays the role of cosmic time.
This proposal indeed can be easily formulated at the level of
action for such null-like background.  Consider a generic action
${\cal L}$ in which a background tachyon field $T$ may has various
forms of interaction with another field $\Phi$ through coupling
$g^I(T)$'s, that is
\begin{equation}
{\cal L}\sim -\fft12(\partial T)^2 + V(T) + g^I(T)f^I(\Phi,
\partial\Phi)+\cdots
\end{equation}
If there exists a solution with ansatz
\begin{eqnarray}
&&ds^2\sim -2g_{uv}dudv + g_{uu}du^2 + \cdots, \\
&&T=T(u), \\
&&\cdots \nonumber
\end{eqnarray}
where there is no $g_{vv}$ term in the metric to insure
$\partial_u$ is a Killing direction.  The tachyon is also assumed
to be a {\sl monotoneous} function of only $u$.  The solution of
$T$ can be reformulated as a constraint in the action through the
Langrangian multiplier,
\begin{equation}
{\cal L}' \sim {\cal L} + \alpha (T-T(u))
\end{equation}
Solving equation of motion for $\alpha$ is equivalent to replacing
$T$ in the action by $T(u)$, and the {\sl on shell} action reads
\begin{equation}
{\cal L}\sim  V(T[u]) + g^I(T[u])f^I(\Phi,
\partial\Phi)+\cdots
\end{equation}
Notice that the kinetic term of tachyon disappears and then $T$
just plays a similar role as {\sl time} variable without any
dynamics. Therefore, it seems that the proposal of null-like
tachyon as cosmic time works quite well with any geometry with a
null-like killing vector.

\acknowledgments This work was supported in part by the Taiwan's
National Science Council under grant NSC95-2811-M-002-013.



\begin{thebibliography}{99}

\bibitem{Maldacena:1997re}
  J.~M.~Maldacena,
  ``The large N limit of superconformal field theories and supergravity,''
  Adv.\ Theor.\ Math.\ Phys.\  {\bf 2}, 231 (1998)
  [Int.\ J.\ Theor.\ Phys.\  {\bf 38}, 1113 (1999)]
  [arXiv:hep-th/9711200].

\bibitem{Witten:1998qj}
  E.~Witten,
  ``Anti-de Sitter space and holography,''
  Adv.\ Theor.\ Math.\ Phys.\  {\bf 2}, 253 (1998)
  [arXiv:hep-th/9802150].




\bibitem{Itzhaki:1998dd}
  N.~Itzhaki, J.~M.~Maldacena, J.~Sonnenschein and S.~Yankielowicz,
  ``Supergravity and the large N limit of theories with sixteen
  supercharges,''
  Phys.\ Rev.\ D {\bf 58}, 046004 (1998)
  [arXiv:hep-th/9802042].

\bibitem{Boonstra:1998mp}
  H.~J.~Boonstra, K.~Skenderis and P.~K.~Townsend,
  ``The domain wall/QFT correspondence,''
  JHEP {\bf 9901}, 003 (1999)
  [arXiv:hep-th/9807137].

\bibitem{Khoury:2001bz}
  J.~Khoury, B.~A.~Ovrut, N.~Seiberg, P.~J.~Steinhardt and N.~Turok,
   ``From big crunch to big bang,''
  %
  Phys.\ Rev.\ D {\bf 65}, 086007 (2002)
  [arXiv:hep-th/0108187].

\bibitem{Craps:2005wd}
  B.~Craps, S.~Sethi and E.~P.~Verlinde,
   ``A matrix big bang,''
  %
  JHEP {\bf 0510}, 005 (2005)
  [arXiv:hep-th/0506180].


\bibitem{Chu:2006pa}
  C.~S.~Chu and P.~M.~Ho,
  ``Time-dependent AdS/CFT duality and null singularity,''
  JHEP {\bf 0604}, 013 (2006)
  [arXiv:hep-th/0602054].

\bibitem{Lin:2006ie}
  F.~L.~Lin and W.~Y.~Wen,
  ``Supersymmteric null-like holographic cosmologies,''
  JHEP {\bf 0605}, 013 (2006)
  arXiv:hep-th/0602124.

\bibitem{Papantonopoulos:2006eg}
  E.~Papantonopoulos, I.~Pappa and V.~Zamarias,
  ``Geometrical tachyon dynamics in the background of a bulk tachyon field,''
  JHEP {\bf 0605}, 038 (2006)
  [arXiv:hep-th/0601152], and references therein.

\bibitem{Polyakov:1998ju}
  A.~M.~Polyakov,
  ``The wall of the cave,''
  Int.\ J.\ Mod.\ Phys.\ A {\bf 14}, 645 (1999)
  [arXiv:hep-th/9809057].


\bibitem{Dixon:1986iz}
  L.~J.~Dixon and J.~A.~Harvey,
  ``String Theories In Ten-Dimensions Without Space-Time Supersymmetry,''
  Nucl.\ Phys.\ B {\bf 274}, 93 (1986).

\bibitem{Seiberg:1986by}
  N.~Seiberg and E.~Witten,
  ``Spin Structures In String Theory,''
  Nucl.\ Phys.\ B {\bf 276}, 272 (1986).




\bibitem{Klebanov}
  I.~R.~Klebanov and A.~A.~Tseytlin,
  ``D-branes and dual gauge theories in type 0 strings,''
  Nucl.\ Phys.\ B {\bf 546}, 155 (1999)
  [arXiv:hep-th/9811035].


  I.~R.~Klebanov and A.~A.~Tseytlin,
  ``Asymptotic freedom and infrared behavior in the type 0 string approach  to
  gauge theory,''
  Nucl.\ Phys.\ B {\bf 547}, 143 (1999)
  [arXiv:hep-th/9812089].

  I.~R.~Klebanov and A.~A.~Tseytlin,
  ``A non-supersymmetric large N CFT from type 0 string theory,''
  JHEP {\bf 9903}, 015 (1999)
  [arXiv:hep-th/9901101].

  I.~R.~Klebanov,
  ``Tachyon stabilization in the AdS/CFT correspondence,''
  Phys.\ Lett.\ B {\bf 466}, 166 (1999)
  [arXiv:hep-th/9906220].


\bibitem{Martelli}
  G.~Ferretti and D.~Martelli,
  ``On the construction of gauge theories from non critical type 0 strings,''
  Adv.\ Theor.\ Math.\ Phys.\  {\bf 3}, 119 (1999)
  [arXiv:hep-th/9811208].


\end{thebibliography}
\end{document}